\documentclass[showpacs,preprint,aps]{revtex4}
\usepackage{graphicx}
\usepackage{amssymb}
\usepackage{amsmath}
\usepackage{yfonts}
\usepackage{mathrsfs}

\begin{document}
\begin{center}
{\bf \Large Inside the Hydrogen Atom}
\vskip 5mm
M. Nowakowski, N. G. Kelkar, D. Bedoya Fierro and A. D. Bermudez Manjarres\\
\vskip 5mm
Departamento de Fisica, 
Universidad de los Andes\\
Cra. 1E, 18A-10, Bogot\'a, Colombia
\end{center}
\begin{abstract}
We apply the non-linear Euler-Heisenberg theory to calculate
the electric field inside the hydrogen atom. We will demonstrate 
that the electric field calculated in the Euler-Heisenberg theory
can be much smaller than the corresponding field emerging from the 
Maxwellian theory. In the hydrogen atom this happens only at
very small distances. This effect reduces the large 
electric field inside the hydrogen atom calculated from
the electromagnetic form-factors via the Maxwell equations.
The energy content of the field is below the pair production threshold. 

\end{abstract}
\maketitle
\section{Introduction}
The electric field extracted from the electromagnetic form-factors of the 
proton is, in the short range 
(where it differs significantly from the Coulomb field), not a correction to the 
latter. This short range part is handled as a perturbation in calculating the energy correction which is possible
due to the rapid fall-off of finite size contribution outside the proton. However, the magnitude of the electric field
reaches a value up to 10$^3$ MeV$^2$ which is not only extremely high, but also 
dangerously close to allowing 
electron-positron pair production. On the other hand, quantum corrections to 
Maxwell's equations known as the
Euler-Heisenberg theory yield a different picture of the same physical phenomenon inside the hydrogen atom: the field
becomes significantly smaller falling below the pair production threshold. We outline the theory which leads to this effect. 

\section{The short range electric field inside the hydrogen atom}
The electromagnetic properties of the nucleons are encoded in their two form-factors,
$F_1$ and $F_2$. In calculating the elastic amplitude of the scattering process, $ep \to ep$,
these form factors enter one of the vertices by making the replacement
\begin{equation} \label{vertex}
e\gamma_{\mu} \to e\left(F_1(q^2)\gamma_{\mu} + \frac{F_2(q^2)}{2m_p}\sigma_{\mu \nu}q^{\nu}\right) \, .
\end{equation}
The standard procedure to extract an interaction potential from the 
elastic amplitude is to (i) expand the latter in terms proportional to powers of
$1/c^2$ (which leads to the non-relativistic amplitude $A_{NR}\equiv V(\mathbf{q})$) and (ii) take the
Fourier transform, i.e, $V(\mathbf{r})=\int d^3q/(2\pi)^3 e^{i\mathbf{q} \cdot \mathbf{r}}V(\mathbf{q})$.
In the case of the electromagnetic interaction, this leads to the Breit equation \cite{Breit, LL}
and if we include the proton form-factors we explicitly take into account in the Breit equation
the finite size of the proton in $V_{ep}(\mathbf{r})$ \cite{us1}. In terms of the so-called
electric ($G_E$) and magnetic ($G_M$) Sachs form-factors
$G_E=F_1 +\frac{q^2}{4m_p}F_2$
and
$G_M = F_1 +F_2$
the result of the scalar interaction potential in momentum space can be written as
\begin{equation} \label{scalar}
V_{ep}({\mathbf{q}})=e\left[\frac{G_E({\mathbf{q}})}{{\mathbf{q}}^2}
- \frac{1}{8m_p^2}G_E({\mathbf{q}}) -\frac{1}{8m_e^2}G_E({\mathbf{q}})\right]\, .
\end{equation}
In principle, the Breit equation gives also the fine and hyperfine structure of the interaction which have to be added
to the term above. We do not spell it out explicitly here since we will not make use of 
it. For details we refer to \cite{us2}. 
We can identify the first term as the Coulomb term (modified due to finite size) $V_C$, the second term is called the proton Darwin 
$V_{pD}$ and the third the
electron Darwin term $V_{eD}$.
We identify the electric potential of the proton from the Breit equation
as terms which are (i) scalar (no spin and no momentum terms which appear in the fine and hyperfine part) 
and (ii) independent from the 
properties of the test particle. By this token the proton Darwin term can be part of
the electric potential, but the electron Darwin not. This has some consequences for the electric field
at small distances inside the proton and also for the charge distribution.
Hence we write 
$V_{ep}({\mathbf{r}})
=eV_C({\mathbf{r}}) + V_{pD}({\mathbf{r}}) + V_{eD}({\mathbf{r}})
=eV_p({\mathbf{r}}) + V_{eD}({\mathbf{r}})$ where $V_p$ is the proton electric potential.
As an example let us take the dipole parametrization 
$G_E({\mathbf{q}}^2)=1/(1+{\mathbf{q}}^2/m^2)^2$. Then  we obtain (see also figure 1)
\begin{equation} \label{dipole}
V_p^C=\frac{e}{r}\left(1-e^{-mr}\left(1+\frac{mr}{2}\right) \right) \, .
\end{equation}
Since the main issues which we want to emphasize will remain mostly insensitive to 
the parametrization of the electric
form-factors, we shall use the dipole parameterization where the Darwin terms can 
also be calculated analytically. The electric field $\mathbf{E}_p$ calculated from 
$V_p^C$ (which includes the form-factors) 
can be now as strong as 10$^3$ MeV$^2$. It is reasonable to ask if this is indeed 
the case. We will give a partial answer 
in the following two sections starting with a discussion of the the non-linear
Euler-Heisenberg theory.
\begin{figure}
\begin{center}
\includegraphics[height=6cm, width=6cm]{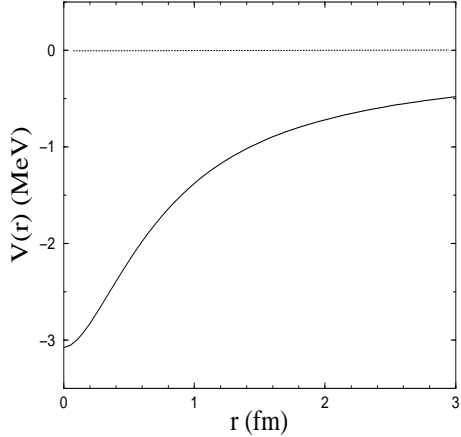}
\end{center}
\label{mo}
\caption{The electric potential $V_p^C$.}
\end{figure}

\section{Static electric field  from Euler-Heisenberg theory}
At the Born level, photons do not interact with each other. However,
the box diagram with four external photon (light-light scattering) legs is 
an example of how
quantum mechanics effectively introduces a four photon vertex.
One can introduce this effect into the Lagrangian language
which is known as the Euler-Heisenberg theory. In accordance with the gauge 
invariance we have to formulate the theory in terms of the two gauge invariant
quantities
 \begin{eqnarray} \label{inv} 
\frac{1}{4}F_{\mu \nu}F^{\mu \nu}&=&
-\frac{1}{2}({\mathbf{E}}^2 -{\mathbf{B}}^2)\,, \nonumber \\
\frac{1}{4}F_{\mu \nu}
\tilde{F}^{\mu \nu}&=&{\mathbf{E}}\cdot {\mathbf{B}}\, .
\end{eqnarray}
Then, we can expect that in the first approximation (weak field approximation)
the new part of the Lagrangian $\delta {\cal L}_{EH}$ is quadratic in these invariants.
Indeed, Euler and Heisenberg obtained \cite{EH}
\begin{equation} \label{EH}
\delta {\cal L}_{EH} = \eta
\left(({\mathbf{E}}^2- {\mathbf{B}}^2)^2 + 7({\mathbf{E}}\cdot 
{\mathbf{B}})^2\right)
\end{equation}
with
\begin{equation} \label{eta}
\eta =\frac{\alpha^2}{360\pi^2 m_e^2}\, .
\end{equation}
Our requirement of 
weak fields means that in
${\cal L} \sim F^2(1+\eta F^2)$
($F \sim E,\,\, B$) the new term must be small as compared to one or
$\eta E^2 \ll 1$ and $ \eta
B^2 \ll 1$. 
It then follows that 
$ E\ll 2 \times 10^3\,\, MeV^2$.
The new inhomogeneous Maxwell's equations now read:
\begin{eqnarray} \label{maxwell}
{\mathbf{\nabla}} \cdot {\mathbf{D}} &=& \rho \, , \nonumber \\
{\mathbf{\nabla}} \times {\mathbf{H}}-\frac{\partial \mathbf{D}}{\partial t}&=&{\mathbf{j}}\, ,
 \end{eqnarray}
with
\begin{eqnarray} \label{P}
{\mathbf{D}} &\equiv& {\mathbf{E}} +4\pi {\mathbf{P}}\, ,   \nonumber \\
{\mathbf{P}} &=& \eta 
\left[4{\mathbf{E}}({\mathbf{E}}^2 -{\mathbf{B}}^2) + 14{\mathbf{B}}({\mathbf{E}}\cdot
{\mathbf{B}})\right] 
\end{eqnarray}
and
\begin{eqnarray} \label{M}
{\mathbf{H}} &\equiv& {\mathbf{B}} -4\pi {\mathbf{M}}   \, ,   \nonumber \\
{\mathbf{M}} &=&   \eta 
\left[-4{\mathbf{B}}({\mathbf{E}}^2 -{\mathbf{B}}^2) + 14{\mathbf{E}}({\mathbf{E}}\cdot
{\mathbf{B}})\right]\, .         
\end{eqnarray}
In the electrostatic case, putting $\mathbf{B}=0$ and neglecting 
all time derivatives we can write
\begin{equation} \label{static}
{\mathbf{\nabla}} \cdot {\mathbf{D}}= \rho= {\mathbf{\nabla}}
\cdot {\mathbf{E}}_0 \, ,
\end{equation}
where $\mathbf{E}_0$ is the electric field as calculated using the 
Maxwellian theory with $\rho$ (we treat
the charge distribution as a given source). 
Hence, assuming spherical symmetry we end up with 
an algebraic equation
\begin{equation} \label{algebraic}
E+ \eta^{\prime} E^3=E_0
\end{equation}
with $\eta^{\prime}= 16 \pi \eta$. The solution is always 
in the form $\mathbf{E}[\mathbf{E}_0]$.
One can extend the above theory to two-loops \cite{2loops} in which case
the algebraic equation will become a quintic polynomial equation of the form
\begin{equation} \label{quintic}
E+\eta_1 E^3 +
\eta_2  E^5=E_0
\end{equation}
where $\eta_1 \simeq \eta^{\prime}$.

\section{The electric field inside the hydrogen atom according to non-linear
electrodynamics}  
We can apply the results of the Euler-Heisenberg theory in the
electrostatic case for $\mathbf{E}_0=\mathbf{E}_p$ where 
$\mathbf{E}_p$ is the electric field calculated using the electromagnetic 
form-factors. The results are shown in figure 2. As 
the electric field calculated from the form-factors is not a correction
to the $1/r^2$ Coulomb field at small distances so is the result from
Euler-Heisenberg not a correction to $E_0$ at small distances. Since
the significant changes happen only at short distances, both modifications
(the finite size and the light-light modifications) will induce only small
corrections to the energy levels.

The energy content of the electric field is given by 
$\mathscr{E}[E]=\int \mathbf{E}^2 d^3x$. For $E=E_p$ we get
\begin{equation} \label{2m1} 
\mathscr{E}[E_p]\simeq 1 \, \, MeV
\end{equation}
which, given the uncertainties in the form-factors, 
is close to the pair production threshold of $2m_e$.
On the other hand we obtain
\begin{equation} \label{2m2}
\mathscr{E}[E_{\gamma \gamma, 1loop}]
\simeq 0.26\, \, MeV
\end{equation}
in the one-loop Euler-Heisenberg theory and
\begin{equation}
\mathscr{E}[E_{\gamma \gamma, 2loop}] 
\simeq 0.076\, \, MeV
\end{equation}
in its two-loops version. Both results are safely below the pair
production threshold.

\begin{figure}
\begin{center}
\includegraphics[height=6cm,width=7cm]{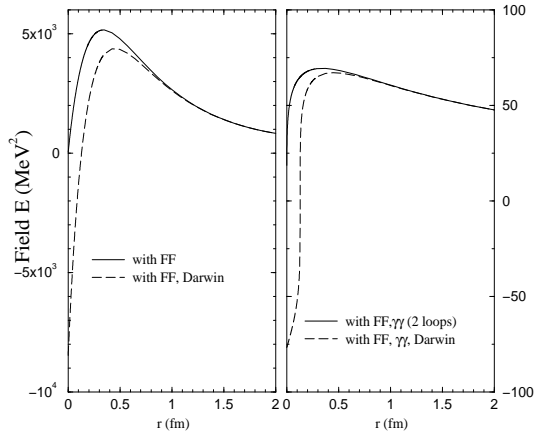}
\end{center}
\label{mo}
\caption{Electric field inside the hydrogen atom. On the left are shown 
the results
from form-factors (FF) with and without the proton Darwin term. On the right
we show the same electric field as emerging from the Euler-Heisenberg theory.}
\end{figure}
\section{Discussion}
Perturbations in the hydrogen atom can be viewed as additional contributions
to the Coulomb potential suppressed by higher powers of the fine structure constant $\alpha$
(for instance the Uehling potential \cite{serber}) or as very short range modifications of the Coulomb
potential. The best paradigm is given by the finite size corrections to the Coulomb part. In the
short range these modifications can not be considered as perturbations of $1/r$, but their short
range character qualifies them as perturbative contributions in the hydrogen atom. A similar situation occurs when
we calculate the full non-linear contributions of light-light scattering. In the short range,
where the non-linear effects are important, the modifications of the Euler-Heisenberg theory
are not perturbative contributions to the Coulomb potential, but as in the case
of the finite size effects they will contribute perturbatively to the observables of the hydrogen atom.
Moreover, the electric field extracted from the electromagnetic form-factors via the Maxwellian theory
is too strong as its energy content is close to the pair production threshold. In contrast to this,
the electric field in the Euler-Heisenberg theory is much smaller in the short range and its energy
comes out to be safely below the pair production threshold. How this affects the electron Darwin term and the
hydrogen observables remains to be explored \cite{future}.

\end{document}